\documentclass[preprint,12pt]{elsarticle}
\usepackage{natbib}

\usepackage[margin=2.25cm]{geometry}
\usepackage{amssymb}
\usepackage[unicode=true]{hyperref}
\usepackage{amsmath}
\usepackage{braket}
\usepackage{xcolor, soul}
\usepackage{booktabs}

\DeclareUnicodeCharacter{2082}{$_2$}
\DeclareUnicodeCharacter{2083}{$_3$}
\journal{International Journal of Hydrogen Energy}

\begin{document}
	
\begin{frontmatter}

\title{A DFT and Machine Learning-Assisted Study on the Lattice Thermal Conductivity of LiCdSb for Thermoelectric Applications} %

\author[inst1,inst2]{R. Zosiamliana}
\author[inst2]{Lalhriat Zuala}
\author[inst3]{N. T. Tien}
\author[inst3]{Vo Khuong Dien\corref{cor2}}
\cortext[cor2]{vokhuongdien@gmail.com}
\author[inst4]{A. Laref}
\author[inst1]{D. P. Rai\corref{cor1}}
\cortext[cor1]{dibyaprakashrai@gmail.com}

\affiliation[inst1]{organization={Advanced Functional Materials \& Simulation Lab (AFMSL), Department of Physics},
	addressline={Mizoram University}, 
	city={Aizawl},
	postcode={796004}, 
	country={India}}
    \affiliation[inst2]{organization={Physical Sciences Research Center (PSRC), Department of Physics},
	addressline={Pachhunga University College}, 
	city={Aizawl},
	postcode={796001}, 
	country={India}}

\affiliation[inst3]{organization={College of Natural Sciences},
	addressline={Can Tho University}, 
	city={Can Tho City},
	postcode={900000}, 
	country={Vietnam}}

 \affiliation[inst4]{organization={Department of Physics and Astronomy, College of Science},
	addressline={King Saud University}, 
	city={Riyadh},
	postcode={11451}, 
	country={Saudi Arabia}}

\date{\today}

\begin{abstract} By using first-principles density functional theory (DFT) and the Boltzmann transport equation, we have calculated the corresponding electronic and thermoelectric properties of LiCdSb. For calculating electron transport properties, accurate band-structure estimation is crucial. Hence, for the precise band gap calculation, we have implemented a hybrid functional HSE06, which is widely known for its high accuracy. To evaluate the thermoelectric performance of a material, the calculation of lattice thermal conductivity (K$_l$) is a key parameter. However, from a theoretical perspective, the calculation of lattice thermal conductivity is very complex and demands huge computational resources. Therefore, in this work, we have opted for an alternative method of machine-learning interatomic potentials (MLIPs) for the calculation of K$_l$. Our result of K$_l$=0.24 Wm$^{-1}$K$^{-1}$ at room temperature is in qualitative agreement with the available theoretical and experimental data. The figure of merit (ZT) with $\kappa_l$ estimated from Slack+TDEC (ZT$^{T=300/500}$=0.18/0.37), and machine learning (ML) models (ZT$^{T=300/500}$=0.17/0.37), combining with HSE06-based electronic transport properties agreed well with available experimental reported value of (ZT$_{exp}^{T=300/500}$=0.10/0.32). However, we report the ZT value well above the benchmark value of $\sim$ 1 beyond 600K. The ZT value exceeding 1 at higher temperatures makes LiCdSb a promising material for high-temperature energy conversion.

\end{abstract}

\begin{keyword}
			DFT \sep Half-Heusler \sep Lattice thermal conductivity \sep MLIP  \sep Slack equation \sep TDEC
\end{keyword}
		
\end{frontmatter}

\section{Introduction}\label{sec:intro}
The discovery of fossil fuels has brought significant advancement in modern-day technologies, revolutionised industrialisation, and improved the living standard of the people.\cite{Rehman2022}. Moreover, the development is sometimes accompanied by environmental disasters. The rapid industrial growth and ambitious technological development demand a huge energy supply. The large energy demand has ultimately created a void in the energy supply--demand mechanism due to the depletion of non-renewable resources (limited fossil fuels)\cite{Xu2023}. In addition, the overutilization of fossil fuels has escalated environmental pollution with the emission of greenhouse gases.\cite{ijerph2018} To address this gap and maintain the balance between life sustainability and economic development, the search for alternative solutions (renewable energy sources) has become a major priority in the 21$^{st}$ century. The main renewable energy sources are solar, wind, hydropower, biomass,  geothermal, ocean and energy from materials manipulation like thermoelectric, piezoelectric, superconductivity, etc.    
\par Among the various energy-harvesting techniques, thermoelectric materials look promising and much greener due to their ability to convert waste heat into electric energy without emitting greenhouse gases. This clean and green approach paved the way for improving energy efficiency and reducing direct dependence on fossil fuels. Thermoelectric devices work on the temperature gradient ($\Delta T$)that creates a potential difference ($\Delta V$) between the two ends and obtain the Seebeck effect ($\frac{\Delta V}{\Delta T}$).\cite{MAMUR2021,FENG2023} As a result, charge carriers diffuse from the hot end to the cold end of the electrode, giving rise to the flow of electric current through an external circuit. Other than their eco-friendly approach, thermoelectric generators (TEGs) offer various practical advantages such as portable design, absence of moving parts, low maintenance, noise-free operation, reliability and size scalability. TEGs have a wide range of applications from micro-level domestic appliances to large macro-scale industrial waste-heat recovery systems. As we know, an enormous amount of heat energy has been wasted from various sectors like industries, automobiles, domestic activities, power plants, etc.; thermoelectric materials provide an innovative solution for energy recovery.\cite{Aridi2021,HUANG2025} However, the development of highly efficient thermoelectric materials by optimizing its parameters for practical application is a major challenge.  So, the primary objective lies in the development of a sustainable energy harvesting system that can effectively solve the global energy crisis without compromising the ecological balance.
\par As reported in earlier studies, HH alloys with low band gaps (0.5--1.8 eV) are suitable for thermoelectric applications\cite{Ioffe,LIU201750}. Ternary half-Heusler (HH) alloys crystallizes in the face-centred cubic structure, having space group (F$4\overline{3}$m \#216). HHs are the most explored materials among the families of Heusler alloys due to a wide range of interesting properties, like non-toxic elemental composition, high melting point, superior electrical conductivity, simple crystal geometry, robust mechanical properties(flexibility in doping), etc. \cite{Yang2008, Li2024} 
HH have shown various characteristics like semiconductor\cite{Corbaci2026,Roy2012,Dubey2024,Yi-Huang2025}, metal/half-metal\cite{Chen2011,Zhang2016}, semimetal\cite{Nakajima, Shekhar}, topological quantum material\cite{Feng2010,Sahni2019}, etc. Semiconducting and metallic behaviour of HHs can be identified based on the Slater-Pauling rule  (M$_t$=Z$_t$-18), where M$_t$ is the total magnetic moment, and Z$_t$ is the total number of valence electrons count (VEC) in the unit cell\cite{Galanakis2002}. For Z$_t$=18, M$_t$=0, which generally represents semiconducting behaviour with a finite band gap, whereas M$_t$ $\not=0$ gives metallic/half-metallic behaviour.  de Groot et al. first predicted the half-metallic ferromagnet in HH NiMnSb with dissimilar characteristics of two spin channels, as one of them was conducting while the other was fully insulating, giving rise to 100\% spin-polarization at the Fermi level.\cite{deGroot} Some of the HH half-metals with M$_t$ $\not=0$ and Z$_t$$\not=18$ are XYZ (X, Y=V, Cr, Mn, Fe, Co and Ni; Z=Al, Ga, In, Si, Ge, Sn, P, As, and Sb)\cite{FENG201492}, XYZ (X=Li, Na, K and Rb; Y=Mg, Ca, Sr and Ba; Z=B, Al and Ga)\cite{UMA2014167},  XCsBa(X = C, Si and Ge),\cite{LAKDJA20138} XCrAl (X=Fe, Co, Ni) and NiCrZ (Z=Al, Ga, In),\cite{LUO2008} XCrZ (X = Li, K, Rb, Cs; Z = S, Se, Te),\cite{Xiaotian} LiMnZ(Z=N,P,Si),\cite{Damewood} VPdZ (Z = Ge, Sn)\cite{KUMAR2025}  etc. While HHs like MNiSn (M = Ti, Hf, Zr)\cite{Yang2008} and MCoSb (M = Ti, Zr, Hf)\cite{Joshi_2019} having Z$_t$=$18$ (VEC) are narrow band gap semiconductors. Yadav et al. studied a large set of Li-based HHs using full-potential linearized augmented plane wave (FPLAPW) within a framework of density functional theory (DFT) and reported variation of band gap from 0.0 eV to 4.23 eV.\cite{Yadav2015} 
\par The thermoelectric efficiency can be estimated from the equation of the figure of merit given by 
\begin{equation}
    ZT=\frac{S^{2}\sigma T}{K_e+K_l}
    \label{zt}
\end{equation}
where $S$, $\sigma$, $T$, $K_e$ and $K_l$ are the Seebeck coefficient, electrical conductivity, absolute temperature, electron part of thermal conductivity and lattice part of thermal conductivity, respectively.\cite{Goldsmid2010} Lattice contribution is a deciding factor in estimating ZT, where $K_l$ plays a crucial role. Since $K_l$ is in the denominator of the figure of merit equation, keeping it low will possibly enhance the ZT value. There are several strategies to reduce $K_l$ via materials modelling, like doping, layering, superlattices, nanostructuring, etc.
\par Over the years, several HHs have been discovered and extensively investigated for potential thermoelectric applications from first-principles theory and experiment. However, the development of HH materials with efficiency close to those of state-of-the-art Pb/Bi-based thermoelectric materials (PbTe and Bi$_2$Te$_3$)\cite{Pin-Wen2005,Wang2024,Hao2016,Caylor2005,Beyer2002} remains a big challenge. Materials modeling and design have gained a significant research interest due to their ability to optimize the material properties at the atomic scale. 
A DFT study using Quantum Espresso software on HfRhSb has reported a band gap of 1.21 eV and ZT value of 0.42 at 1200 K.\cite{Kaur2018} Kaur et al., reported an improvement in the band gap of HfPtPb from 0.58 eV to 0.73 eV while using TB-mBJ over GGA and a low lattice thermal conductivity of 9.9 W/mK with maximum ZT=0.25 at 1000 K.\cite{Kaur2017} Yadav et al., on assuming the relaxation time 10$^{-14}$ s have found the maximum thermoelectric power factor of 38.46 ($\mu$W/cm K$^2$s) for  n-type LiZnSb comparable to that of PbTe (35 $\mu$W/cm K$^2$s) at 300 K.\cite{Yadav2015} 
Guo et al. investigated the electronic and thermoelectric properties of ZrNiPb using GGA and GGA+SOC and reported a band gap of 0.37 eV, a lattice thermal conductivity of 14.5 W m$^{-1}$K$^{-1}$ and ZT=0.30 at room temperature.\cite{Guo2016} 
Anuradha et al., have studied the Li-based HHs LiMgX(X = N, P, As) and reported a large power factor of 2.31$\times$10$^{12}$ W m$^{-1}$s$^{-1}$K$^{-2}$ at 1200 for LiMgN.\cite{Anuradha} 
Wei et al. performed a ﬁrst-principles simulation and semi-classical Boltzmann transport theory to investigate the electronic, optical and thermoelectric properties of TaCoSn and reported narrow-gap behaviour with the highest power factor of 10.93 $\mu$cm$^{-1}$K$^{-2}$ at 820 K.\cite{Wei2018}. S. A. Khandy investigated the electronic and thermoelectric properties of the six HH alloys, and among them, PtTaAl, PtTaGa and PtTaIn are reported to be the potential thermoelectric materials having power factor up to 90.5, 106.7, 106.5 mW/(K$^2$m), respectively, at 900K.\cite{Khandy2021} Jaishi et al., from ab-initio calculation reported the p-type indirect semiconducting band gap of 0.94--1.01 eV for Rhodium-based HHs and maximum ZT$\sim$1 for RhTiBi, whereas for RhTiP, RhTiAs, and RhTiSb ZT lies between 0.38 and 0.67.\cite{Jaishi2022}

\par There are several reports on the enhancement of ZT above 1 by materials modeling and structural modulation via doping. A recent thorough first-principles calculation on the high-entropy half-Heusler alloy Li$_{0.5}$Na$_{0.5}$CuSe$_{0.5}$Te$_{0.5}$ reported ZT value of 1.005 at room temperature.\cite{Zhang2025} Ghosh et al. have performed a high-entropy-driven study on the thermoelectric properties and reported a record-breaking ZT value of 1.5 at 1160K in  Nb$_{0.25}$Ta$_{0.25}$Ti$_{0.25}$V$_{0.25}$ among the HHs.\cite{Ghosh2024} By implementing a co-doping (Ta and Ti) strategy, the ZT value of (Nb$_{0.6}$Ta$_{0.4}$)$_{0.8}$Ti$_{0.2}$FeSb has been enhanced to 1.6 at 1200K with suppressed lattice thermal conductivity.\cite{Junjie2018} Li et al. performed a phase boundary mapping experiment and obtained the highest ZT value 0.71 at 973 K for ZrNi$_{1.11}$Sn$_{1.04}$\cite{Li2020}. Another experimental work on Nb$_{0.55}$Ta$_{0.40}$Ti$_{0.05}$FeSb have reported a maximum power factor of 78$\mu$Wcm$^{-1}$K$^{-2}$ approaching the ZT value $\sim$ 0.86 at 300-873 K.\cite{Poudel} A first-principles study on the doped material Zr$_x$Hf$_{1-x-y}$Ta$_y$NiSn has reported enhancement in the ZT value from 0.55 to 0.75 with 50\% Ta-doping.\cite{Rai2015} Huang et al., from their experiment, reported a low lattice thermal conductivity of $\sim$3.6 Wm$^{-1}$K$^{-1}$ at 340 K and a highest ZT value of 0.45 at 823 K for Yb$_{0.95}$Ta$_{0.05}$NiSb.\cite{Huang2022} Zhu et al. synthesized the TaFeSb-based HH, creating an alloying structure of Ta$_{0.74}$V$_{0.1}$Ti$_{0.16}$FeSb, exhibiting a ZT value of $\sim$1.52 at 973 K.\cite{Zhu2019}
Mitra et al. have employed the alloying technique and achieved a ZT value in the range of 0.8--1.5 for Hf$_{0.3}$Zr$_{0.7}$CoSn$_{0.3}$Sb$_{0.7}$ with the low lattice thermal conductivity (2.5--4.0)W/Km.\cite{Mitra2022}
Moreover, engineering materials behaviour is a complex task and requires extensive efforts. An alternative and simple technique is data mining via machine learning (ML) algorithms to sort out the HH materials having ultralow lattice thermal conductivities.\cite{Han2021,Hidetoshi2021,Athar2026,Florenciano,Yuqing2026,ZHANG2026}  

Dylla et al., by constructing an “orbital phase diagram" as the base, have used the ML tool to find the semiconductor HHs having W-pocket valence band maxima favourable for thermoelectric applications.\cite{Dylla2020}
Zhu et al. have successfully integrated graph neural networks and random forest algorithms into an ML model to obtain low lattice thermal conductivity and identified a set of new rare-earth chalcogenides-based thermoelectric materials having ZT  as low as $\sim$1.1 at 800 K.\cite{Zhu2021} 
Tranas et al. implemented a non-linear ML model incorporating the random forest algorithm on 122 HH samples and identified low lattice thermal conductivity in the range of 0.85 to 23.45 (W/mK), in which LaPtSb exhibit the lowest, and LiBSi is the highest.\cite{Tranas2022} Legrain et al., using a similar ML-based Scikit-learn package\cite{scikit}, have screened out 15 most stable thermoelectric Half-Heuslers among the 323 compositions.\cite{Legrain2017} Another work from the Scikit-learn package using an unsupervised ML model has screened the best 61 HHs out of 456 samples and reported the lowest lattice thermal conductivity of $\sim$4.2 W/mK in the p-type Sc$_{0.7}$Y$_{0.3}$NiSb$_{0.96}$Sn$_{0.04}$ among all of them.\cite{Xue2022}
Filanovich et al. have generated a large number of possible samples of double half-Heuslers from four different elemental compositions and studied their thermal and elastic properties by using the gradient boosting ML model. In this study, they have reported the lowest room temperature value of K$_l$ $\sim$3.69 (W/mK) for MgZrNi$_2$Bi$_2$.\cite{Filanovich2023}  
Yang et al. applied the Machine-learning interatomic potential (MLIP) on 31891 samples to investigate the eﬀect of 4ph scattering and the number of valence electrons on the K$_L$ of HHs. They have reported 80 HHs having K$_l$ values in the range of 0.44--33.16 W/mK at 300 K, in which 8-VEC HHs exhibit lower K$_l$ than 18--VEC HHs owing to a large phonon scattering rate and smaller 2$^{nd}$-order inter-atomic force constants.\cite{Yang2024}
Ojha et al. aim to predict key parameters like the energy band gap for thermoelectric applications using crystal graph convolutional neural networks (CGCNN) ML techniques and validated experiments. This ML model has predicted the band gap of 0.16--0.38 eV, which is in close alignment with the experimental results of 0.203--0.39 eV.\cite{Ojha2025}  The stacking ML model integrating random forest and XGBoost algorithms has given the highest ZT=1.19 and lowest K$_l$=4.2 W/mK in ErNiBi among all the studied HHs.\cite{Kurian2025} Fronzi et al. have performed the ML on a 15,000-sample dataset for the prediction of potential thermoelectric materials within the Kolmogorov–Arnold networks formalism. In this study, they reported a mean absolute error of 34.44 $\mu$V/K and 0.065 eV for the Seebeck coefficient and band gap, respectively.\cite{Fronzi2026}

The efficient data analysis, its execution within a feasible time-frame, and the qualitative agreement of the results, especially in predicting the lattice thermal conductivities, have highlighted the power of machine learning approaches for discovering high-performance thermoelectric materials. In thermoelectrics,  the calculation of K$_l$ is very complex and demands huge computational resources. This is because it needs to determine the harmonic phonons along with the anharmonic phonon-phonon interactions within the first Brillouin zone. As a result, a large number of second- and third-order interatomic force constants (IFCs) are generated by infinitesimal displacements, creating a large number of supercells. 
The anharmonic force constants are evaluated, incorporating the huge number of phonon-scattering processes involved in the Boltzmann transport equation. Consequently, there is an abrupt rise in the computational cost while solving these equations self-consistently, making it a mammoth task to obtain the solutions via DFT. Hence, we have implemented an alternative MLIP method for the prediction of the lattice thermal conductivity, which is computationally very rigorous from DFT/DFPT. Dien et al. have already validated the accuracy and computational reliability of MLIP on estimating the thermoelectric properties of penta--InP$_5$.\cite{Dien2025}

\section{Computational Detail}
\subsection{Parameters during DFT calculation}
\label{Computational Detail (a)}
The first principles work performed in this study is based on the standard-DFT as programmed in Vienna \textit{Ab} \textit{initio} Simulation Package (VASP) version 6.5.1, which rely on the projector augmented wave (PAW) method constituting the core ionic interaction \cite{Hafner2008b,Kresse1996f}. The electron correlation exchange energy within the generalized gradient approximation (GGA) as proposed by Perdew-Burke-Ernzerhof (PBE) formalism was utilized to treat the electron interactions \cite{Perdew1996m}. During the structural optimization, a full-cell optimization without imposing any constraint was firstly conducted and based on Birch-Murnaghan's equation of states (EOSs) \cite{Murnaghan1944d} the ground state lattice was achieved by optimizing the lattice parameter. An energy convergence criteria of 10$^{-6}$ eV was sampled such that forces on each atom was lower than 0.01 eV/{\AA} \textit{via} conjugate gradient (CG) algorithm \cite{Nazareth2009} was adopted to ensure convergence of cell energies and structural parameters. An energy cut-off of 400 eV was considered for plane-wave expansion of the electronic wave function. Within the Monkhorst-Pack grid \cite{Monkhorst1976l}, an explicit k-mesh of 10$\times$10$\times$10 was used to integrate the first Brillouin zone using a Gaussian smearing method of smearing width 0.05 eV. For properties calculations, similar convergence criteria was set-up with a much denser k-mesh of 12$\times$12$\times$12. Also, apart from PBE-GGA approximation, a spin-orbit coupling (SOC) and a hybrid-DFT approach employing HSE06 functional \cite{Heyd2003d} was implemented to achieve more accurate electron interaction treatment. Herein, to determine the TDEC, the Thermo\_PW package coupled with an open-source code Quantum Espresso (QE) version 7.5 was employed\cite{Giannozzi2020d}. The theory of TEDC has been thoroughly reported in \cite{Zhang2021a,Malica2022,Gong2025}.
   
\subsection{ML assisted lattice thermal conductivity prediction}
\label{Computational Detail (b)}
To evade rigorous procedure to calculate the K$_L$ we have opted an alternative approach of Machine Learning (ML) modeling. At first, on-the-fly machine learning potentials (FMLP) model was trained using the machine learning module. To generate a reliable dataset for training the FMLP model, a reliable dataset were generated from an $ab$-$initio$ molecular dynamics (AIMD) simulations at 300 K \cite{Iftimie2005}. A canonical ensemble is used with 10000 steps, each with a time step of 1fs, to obtain configurations close to equilibrium. During the training process, the weights for energy, forces, and stresses are assigned values 1, 0.1, and 0.001, respectively. The second- and third-order interatomic force constants (IFCs) are calculated using the FMLP method in combination with the Phonopy and Phono3py codes \cite{Togo2022,Togo2023} employing a supercell size of 4$\times$4$\times$4. The lattice thermal conductivity is computed using the following equation:

\begin{equation}
	\kappa_{l}^{\alpha\beta}
	=
	\frac{1}{k_{\mathrm B} T^{2}\Omega N}
	\sum_{\lambda}
	f_{0}(\omega_{\lambda})
	\left[f_{0}(\omega_{\lambda})+1\right]
	(\hbar\omega_{\lambda})^{2}
	v_{\lambda}^{\alpha}
	v_{\lambda}^{\beta}
	\tau_{\lambda}
	\label{eq1}
\end{equation}
In equation \ref{eq1}, $k_{\mathrm{B}}$, $T$, $\Omega$, $\hbar$, and $N$ denote the Boltzmann constant, absolute temperature, unit-cell volume, reduced Planck constant, and the total number of sampled wave vectors points in the first Brillouin zone, respectively. The quantities $\omega_{\lambda}$, $v_{\lambda}^{\alpha}$, $v_{\lambda}^{\beta}$, and $\tau_{\lambda}$ represent the phonon angular frequency, group velocity components along the $\alpha$ and $\beta$ directions, and relaxation time of phonon mode $\lambda$, respectively, while $f_{0}(\omega_{\lambda})$ denotes the Bose-Einstein distribution function. To ensure convergence of the lattice thermal conductivity calculations, a $q$-point mesh of $30 \times 30 \times 30$ was employed within the Phono3py framework.
\section{Results and Discussions}
\subsection{Structural Stability and Electronic Properties}
\label{Structure}
\begin{table}[hbt!]
	\small
	\centering
	\caption{\ Wyckoff atomic position of LiCdSb for $\alpha$-, $\beta$-, and $\gamma$-phases.}
	\label{tab1}\renewcommand{\arraystretch}{1.55}
	\begin{tabular*}{0.4\textwidth}{@{\extracolsep{\fill}}llll}
		\hline
		Atoms &$\alpha$-phase&$\beta$-phase& $\gamma$-phase\\ \hline
		Li &4c(y, y, y)&4b(z, z, z)&4a(x, x, x)\\ 
		Cd &4b(z, z, z)&4a(x, x, x)&4c(y, y, y)\\
		Sb &4a(x, x, x)&4c(y, y, y)&4b(z, z, z)\\ 
		\hline
	\end{tabular*}
 \par Here, x=0.00, y=0.25, and z=0.50.
\end{table}
\begin{figure*}[htb!]
	\centering
	\includegraphics[width=17.5cm]{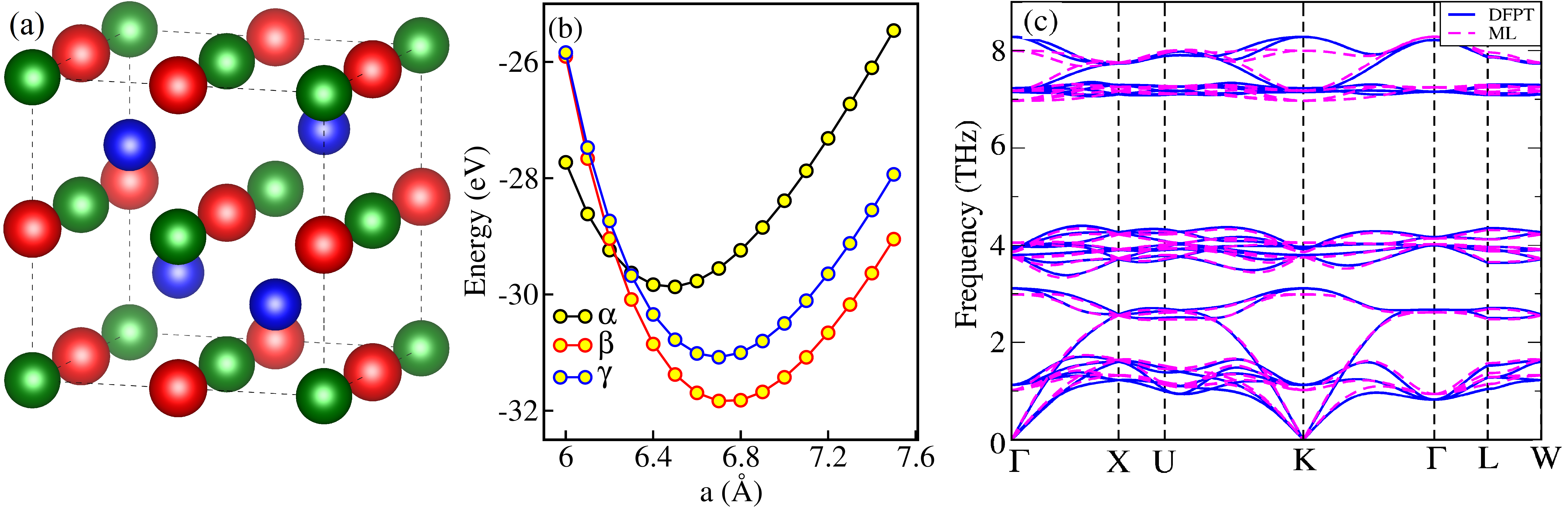}
	\caption{\label{fig1} (a) Optimized structure of $\beta$-phase LCS, where red, green, and blue spheres represent Li, Cd, and Sb atoms, (b) Energy variation vs lattice constant of $\alpha$-, $\beta$-, and $\gamma$-phases using Birch-Murnaghan curve fitting method, and (c) Calculated phonon dispersion curve for $\beta$-phase LCS HH compound.}
\end{figure*}

\par Structural optimization shows that the LCS HH compound crystallizes in the cubic C1$_b$-type structure with space group F-43$m$ (No. 216), corresponding to the MgAgAs-prototype \cite{Yin2015}. This structure admits three distinct atomic arrangements, referred to as the $\alpha$-, $\beta$-, and $\gamma$-phases. The Wyckoff positions occupied by the three in-equivalent atoms in these configurations are listed in table \ref{tab1} for clarity. Most stable configuration was assessed through total energy versus lattice constant plot, shown in figure \ref{fig1}(b), based on the Birch-Murnaghan's equation of state (EOS) (refer equation \ref{eq2}). The smooth parabolic curves observed suggest $\beta$-phase as ground-state configuration, in agreement with previous theoretical studies. The optimized lattice constant value of 6.73 {\AA} compares well with earlier reported values in the range of 6.55-6.74 {\AA}, calculated within the LDA, GGA-PBE, and WC-GGA functionals by Pallavi \textit{et al.,} Bouhemadou \textit{et al.,} and Yang \textit{et al.,} respectively \cite{Pallavi2023, Yang2022,Bouhemadou2015}.

\begin{equation}
	\begin{split}
		E(V)=E{_0}+\frac{9{\times}B{_0}V{_0}}{16} [{ [(\frac{V_0}{V})^\frac{2}{3}-1]{^3}{\times}B'{_0}}+[(\frac{V_0}{v})^\frac{2}{3}\\
		-1]^2{\times}[6-4\times(\frac{V_0}{V})^\frac{2}{3}]]
	\end{split}
	\label{eq2}
\end{equation}

\begin{figure*}[tb]
	\centering
	\includegraphics[width=0.45\textwidth]{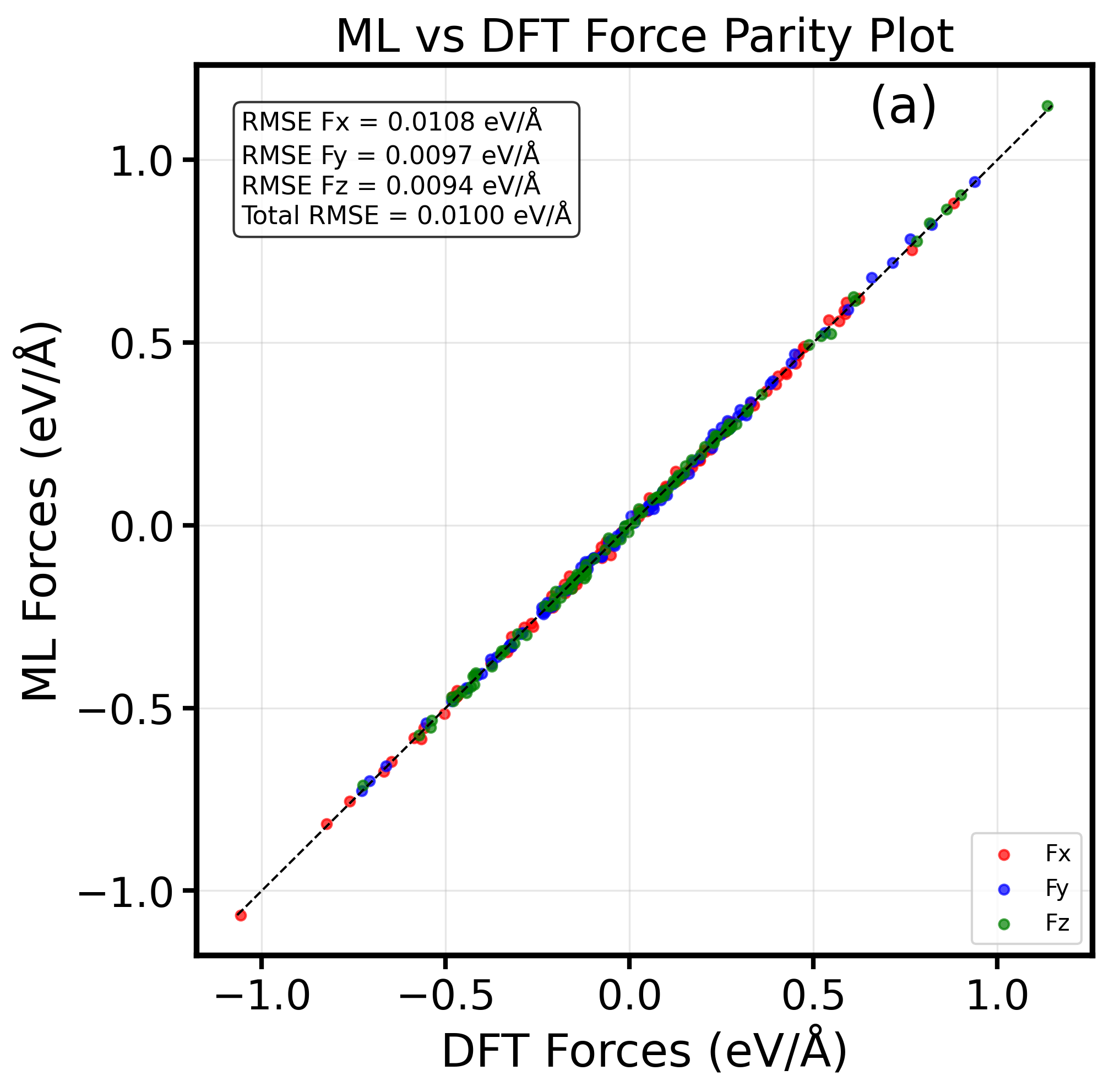}
	\includegraphics[width=0.45\textwidth]{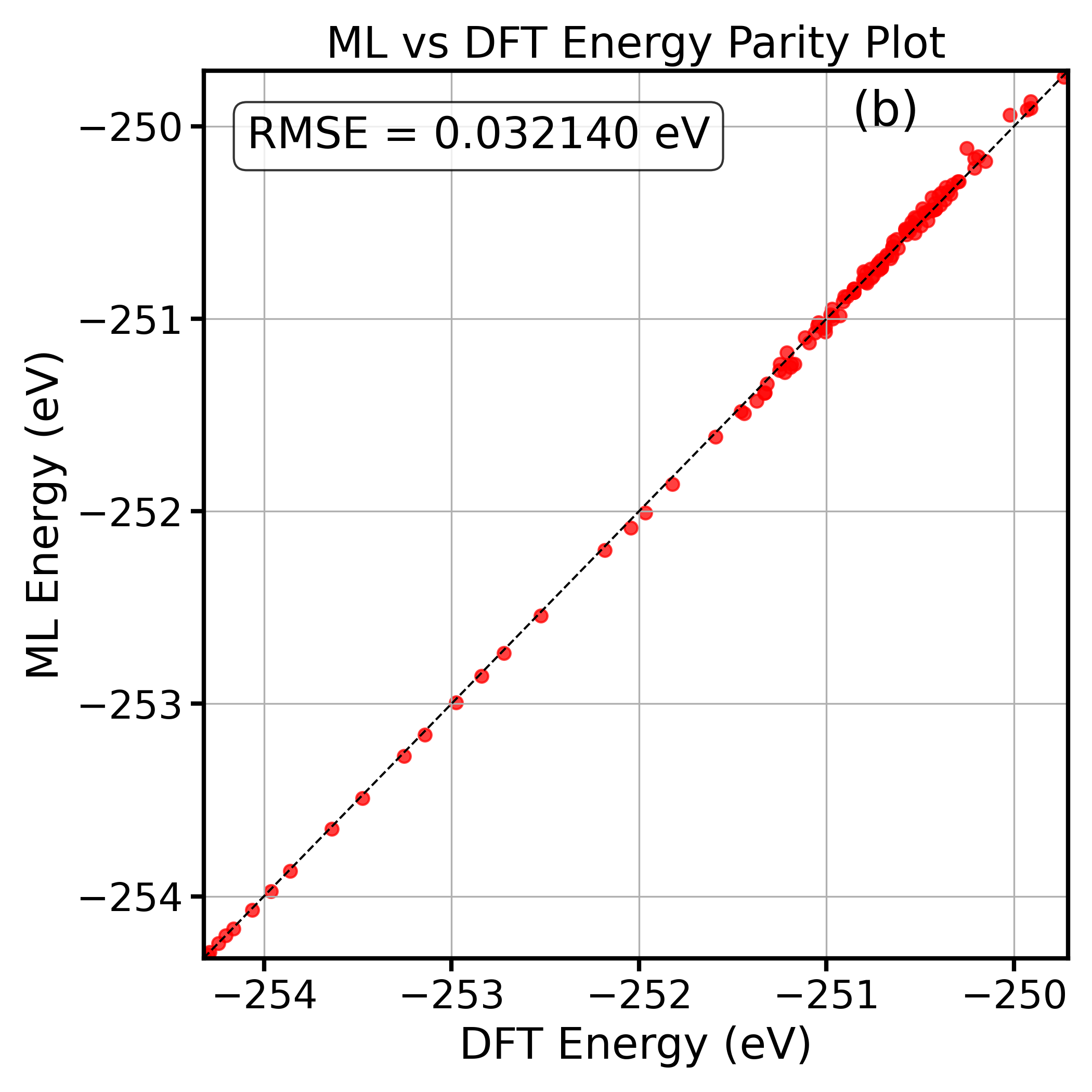}
	\caption{(a) Forces and (b) Energy calculated from DFT and ML.}
	\label{fig2}
\end{figure*}

\par The dynamical stability was examined from the phonon dispersion relations calculated using both density functional perturbation theory (DFPT) and the machine-learning (ML) approach. As shown in Figure \ref{fig1}(c), the absence of imaginary phonon frequencies throughout the Brillouin zone confirms the dynamical stability of the compound. Moreover, the ML-predicted phonon branches closely reproduce the DFPT results over the entire frequency range, exhibiting only marginal deviations near the higher-frequency optical modes. The overall agreement between the two approaches demonstrates that the trained ML model reliably captures the interatomic force constants and lattice vibrational characteristics of LCS, thereby providing an efficient and accurate alternative to computationally demanding first-principles phonon calculations. The predictive capability of the trained ML interatomic potential (MLIP) was evaluated through parity analyses of the atomic forces and total energies with respect to the corresponding DFT data, shown in Figure \ref{fig2}(a) and (b). The force parity plot demonstrates excellent linear correlation between MLIP-predicted force components and the first-principles results. The root-mean-square errors (RMSEs) for the Cartesian components are exceptionally low, yielding 0.0108 eV/{\AA} for F$_x$, 0.0097 eV/{\AA} for F$_y$, and 0.0094 eV/{\AA} for F$_z$, which culminate in a total force RMSE of 0.0100 eV/{\AA}. Concurrently, energy parity analysis displays the agreement between the MLIP and the DFT total energies, manifesting a minimal RMSE of 0.0321 eV. The tight clustering of data points in both parity analyses represents the robustness of the trained potential. These benchmarks suggest that the ML model enables high-fidelity predictions of lattice dynamical properties at a fraction of the computational cost of conventional first-principles methods.    
 
\begin{figure*}[htb!]
	\centering
	\includegraphics[width=13cm]{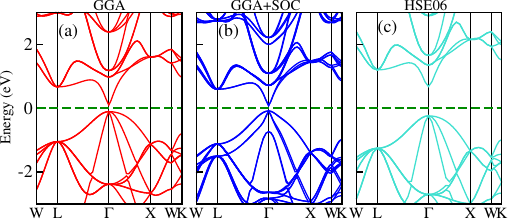}
	\caption{\label{band} Calculated electronic band structure using three different formalism: (a) GGA, (b) GGA+SOC, and (c) HSE06. The green dashes line at 0 eV represents $E_F$.}
\end{figure*}
\begin{figure}[t!]
	\centering
	\includegraphics[width=6.5cm]{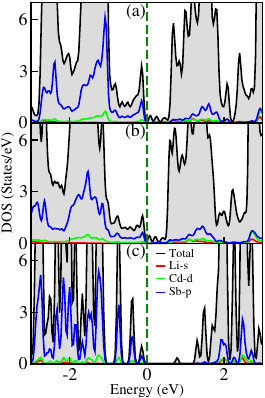}
	\caption{\label{atomic-dos} Calculated total density of state (TDOS) and partial density of state (PDOS) using: (a) GGA, (b) GGA+SOC, and (c) HSE06.}
\end{figure}
\begin{table*}[hbt!]
	\small
	\centering
	\caption{\ Calculated charge transfer from Bader charge analysis (Q$^t$) in $\vert$e$\vert$, the electronic band gap (E$_g$) in eV, both from GGA, GGA+SOC and HSE06, and estimated carrier effective masses m$^*$ in electron rest mass unit m$_0$.}
	\label{table1}\renewcommand{\arraystretch}{1.5}\setlength{\tabcolsep}{1.5pt}
	\begin{tabular*}{0.9 \textwidth}{@{\extracolsep{\fill}}cccc|ccccccc} 
		\toprule
		   & Q$^t$(GGA) & Q$^t$(GGA+SOC) & Q$^t$(HSE06) & E$_g$(GGA) & E$_g$(GGA+SOC) & E$_g$(HSE06) & $\vert$m$^*_e$$\vert$ & $\vert$m$^*_{lh}$$\vert$ & $\vert$m$^*_{hh}$$\vert$&$\vert$m$^*_{d}$$\vert$\\ \hline 
		Li &0.866&0.867&0.875&&&&&&&\\ 
		Cd &0.169&0.166&0.311&0.21&0.002&0.92&0.11&0.08&1.91&1.92\\
		Sb &-1.035&-1.033&-1.186&&&&&&&\\
		\hline
	\end{tabular*}
\end{table*}
To analyze the electronic properties of the studied LCS compound, we perform calculations of band structure, density of states (DOS), and Bader charge analysis \cite{Henkelman2006e}, employing the GGA, GGA+SOC and HSE06.
\par Since the investigated LCS HH alloy crystallized in C1$_b$, the magnetic behaviour can be anticipated using the Slater-Pauling rule \cite{Fecher2006,Galanakis2023}, which predicts the total magnetic moment as m$_t$=z$_t$-18, where m$_t$ and z$_t$ are the total magnetic moment per formula unit and the total number of valence electrons. For the present compound, Li, Cd and Sb contribute 1, 12, and 5 valence electrons, respectively, yielding z$_t$=18. Consequently, the predicted m$_t$ is 0 $\mu_B$. To verify this prediction, spin-polarized DFT (SP-DFT) calculation was performed within the GGA framework, which converged to a non-magnetic ground state with a vanishing total magnetic moment, confirming the semiconducting and non-magnetic character of this compound. Therefore, subsequent GGA+SOC and HSE06 calculations were carried out without imposing any initial atomic spin polarization. 
\par The electronic band gap (E$_g$) tabulated in Table \ref{table1} and the band structure presented in Figure \ref{band}, signify the direct semiconducting nature of the LCS compound, characterized by an electronic transition at $\Gamma$ point, irrespective of the exchange-correlation functional employed. From the DOS plot shown in Figure \ref{atomic-dos}, the upper valence bands are predominantly composed of Sb-p states with appreciable hybridization with Cd-d orbitals, whereas the conduction band edge is primarily derived from the anti-bonding Cd-d/Sb-p states, with a minor contribution from Li-s orbitals. This highlights the dominant role of Cd-Sb covalent hybridization in governing the low-energy gap electronic structure. Within the GGA approximation, the E$_g$ is underestimated due to an intrinsic self-interaction error. While the incorporation of SOC induces relativistic splitting of the Sb-p derived states and modifies the Cd-d/Sb-p hybridization, leading to a slight upward shift of the VBM and a corresponding reduction in the E$_g$. In contrast, the HSE06 hybrid functional substantially enhances the exchange interaction through screened Hartree-Fock exchange, thereby increasing the energy separation between the occupied Sb-p and unoccupied Cd-d derived states, resulting in a significantly enlarged E$_g$. These results demonstrate that the band-edge positions of the electronic structure are primarily dictated by Cd-Sb orbital hybridization, while the competing effects of relativistic interactions and hybrid exchange determine the quantitative value of the E$_g$. 
\par The Bader charge analysis in Table \ref{table1} consistently indicates electron transfer from Li and Cd toward Sb, confirming the predominantly ionic Li-Sb interaction accompanied by appreciable Cd-Sb covalent hybridization. The relatively small positive charge on Cd compared with Li suggests that Cd participates mainly in directional covalent bonding through strong hybridization between Cd-d and Sb-p orbitals, consistent with the PDOS plot. The inclusion of SOC produces only marginal changes in the Bader charges, indicating that relativistic effects primarily modify the band-edge electronic states without significantly altering the ground-state charge distribution. In contrast, the HSE06 functional increases the electron transfer from both Li and Cd to Sb, particularly enhancing the Cd charge from 0.169 to 0.311 $\vert$e$\vert$, indicative of a strengthened electronic localization and a more accurate description of the Cd-Sb bonding interaction. Also, the calculated carrier effective masses are consistent with the band structure shown in Figure \ref{band}(c). The highly dispersive conduction band around the $\Gamma$-point gives rise to a low electron effective mass ($\vert$m$^*_e$$\vert$=0.11m$_0$), indicating high electron mobility. Conversely, the valence-band maximum comprises multiple bands with markedly different curvatures, yielding a broad distribution of hole effective masses ($\vert$m$^*_h$$\vert$=0.08-1.91m$_0$). This anisotropic band dispersion originates from the hybridized Sb-p and Cd-d states near the valence-band edge, resulting in the coexistence of light (m$^*_{lh}$) and heavy holes (m$^*_{hh}$) that is favorable for carrier transport. Since, both the m$^*_{lh}$ and m$^*_{hh}$ contributed to the hole transport, hole DOS effective mass ($\vert$m$^*_d$$\vert$) is evaluated using the formula: $\vert$m$^*_d$$\vert$=$\big[(\vert m^*_{lh} \vert)^\frac{3}{2}+(\vert m^*_{hh} \vert)^\frac{3}{2}\big]^\frac{2}{3}$. Here, $\vert$m$^*_d$$\vert$ is closely equal to $\vert$m$^*_{hh}$$\vert$, this reveal that hole transport is dominated by m$^*_{hh}$. 
            
\subsection{Temperature Dependent Elastic Properties (TDEC) \& Thermodynamic Properties}
\label{Elastic}
\begin{figure}[htb!]
	\centering
	\includegraphics[width=8.5cm]{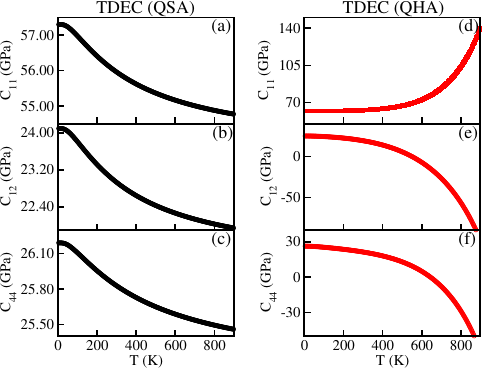}
	\caption{\label{Cij} Calculated TDEC (C$_{ij}$(T)) based on QSA and QHA approach.}
\end{figure}
\begin{figure}[htb!]
	\centering
	\includegraphics[width=8.5cm]{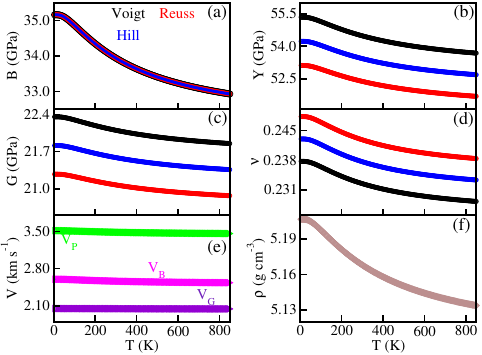}
	\caption{\label{moduli-QSA} Calculated temperature-dependent elastic moduli (TDEM) and other parameters based on the QSA model.}
\end{figure}
The ground state elastic properties of LCS compound was determined employing the finite stress-strain formalism as implemented in the thermo\_pw package. The estimated elastic constants (C$_{ij}$), elastic moduli and other mechanical parameters derived from this approach were listed in Table \ref{table2}. Since, the LCS compound existed in cubic phase, the elastic tensor is completely characterised by only three independent elastic constants, namely C$_{11}$, C$_{12}$, and C$_{44}$, corresponding to the longitudinal, coupling, and shear responses, respectively. The mechanical stability was examined using the Born-Huang stability criteria \cite{Mouhat2014k}. The calculated second-order elastic constants fulfill all the necessary stability conditions (equation \ref{3}), confirming the intrinsic mechanical stability of the LCS compound against infinitesimal lattice deformations.  
\begin{equation}
	C_{11}>0, C_{44}>0, C_{11}-C_{12}>0, C_{11}+2C_{12}>0
	\label{3}
\end{equation} 
\begin{table*}[!]
	\small
	\centering
	\caption{\ Calculated elastic constants C$_{ij}$, elastic moduli such as bulk modulus (B), Young's modulus (Y) and shear modulus (G), all in GPa unit. And other mechanical parameters including Poisson's ratio '$\nu$' (unitless), Pugh's ratio 'k' (unitless), sound velocities 'V' (in km s$^{-1}$ ), Kleinman coefficient '$\zeta$' (unitless), anisotropic factor 'A$_{an}$' (unitless), melting temperature 'T$_m$' (in K), and Debye temperature '$\Theta_D$' (in K). Here, the reported values are calculated based on the stress-strain method.}
	\label{table2}\renewcommand{\arraystretch}{1.5}\setlength{\tabcolsep}{1.5pt}
	\begin{tabular*}{0.9\textwidth}{@{\extracolsep{\fill}}cccccccccccc} 
		\toprule
		C$_{11}$&C$_{12}$&C$_{44}$&B$_v$&B$_r$&B$_h$&Y$_v$&Y$_r$&Y$_h$&G$_v$&G$_r$&G$_h$\\
		55.61&22.26&26.38&33.37&33.37&33.37&55.11&52.89&53.99&22.49&21.39&21.95\\ \hline
		$\nu_v$&$\nu_r$&$\nu_h$&k&V$_P$&V$_G$&V$_B$&V$_{av}$&$\zeta$&A$_{an}$&T$_m$$\pm$300&$\Theta_D$\\
		0.22&0.24&0.23&1.52&3.46&2.05&2.53&2.63&0.68&1.58&881.66&228.44\\
		\hline
	\end{tabular*}
\end{table*}
\begin{figure*}[t!]
	\centering
	\includegraphics[width=12cm]{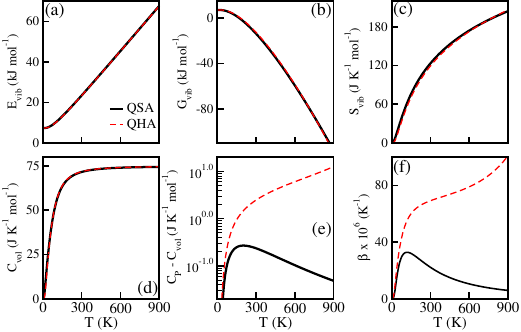}
	\caption{\label{dynamic1} Calculated temperature-dependent thermodynamic properties (TDTP) using QSA and QHA models.}
\end{figure*}
\begin{figure*}[htb!]
	\centering
	\includegraphics[width=12cm]{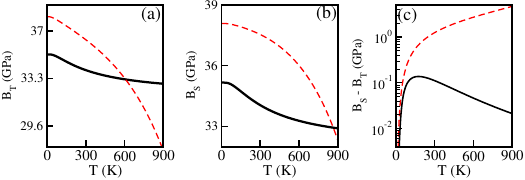}
	\caption{\label{dynamic2} Calculated temperature-dependent bulk modulus (TDBM) using QSA and QHA models. Here, B$_S$ and B$_T$ are the isothermal and adiabatic bulk moduli.}
\end{figure*}
\par From Table \ref{table2}, the C$_{11}$ $>>$ C$_{44}$ indicates the LCS compound is more resistant to axial compression than shear deformation, which is re-evaluated by its bulk modulus (B) greater than shear modulus (G). Here, the elastic moduli: B, Y and G are represented in terms of Voigt, Reuss, and Hill assumptions, which rely on uniform strain, uniform stress, and the average thereof, respectively \cite{Reuss1929h,Hill1952h}. The relatively low values of C$_{11}$, C$_{12}$, and C$_{44}$, together with the corresponding B, Y, and G, compared to conventional full- and quaternary-Heusler alloys, indicate a mechanically compliant lattice, which is characteristic of the open C1$_b$ half-Heusler structure \cite{Sanga2026,Zosiamliana2023c}. Also, the close agreement between the Voigt, Reuss, and Hill averages demonstrates weak elastic anisotropy in the polycrystalline response. Furthermore, $\nu_H$=0.23 and k=1.52 suggest a predominantly brittle character with appreciable covalent bonding. The $\zeta$=0.68 suggests that bond-angle bending is energetically favoured over bond stretching under external deformation \cite{Kleinman1962i}, while the A$_{an}$=1.58 indicates mediocre elastic anisotropy \cite{Sun2005d}. Moreover, the relatively low acoustic velocities are consistent with the small elastic stiffness and give rise to a moderate $\Theta_D$ (= 228.44 K), indicative of relatively soft lattice vibrations. The estimated T$_m$ of 881.66 $\pm$ 300 K further suggests adequate thermal stability, highlighting the suitability of LCS for thermoelectric application under moderate operating temperatures.
\par In this study, the C$_{ij}$(T) are evaluated within both the quasi-static (QSA) and quasi-harmonic approximations (QHA) \cite{Wang2010,Keuter2019,Winey2020,Malica2021c}. The two approaches yield distinct temperature-dependent trends, shown in Figure \ref{Cij}, mainly due to their fundamentally different treatments of thermal effects. Within the QSA, the C$_{ij}$(T) originates exclusively from thermal expansion through the equilibrium volume, while the intrinsic vibrational contribution to the elastic response at a fixed volume is neglected. In contrast, the QHA explicitly incorporates lattice vibrational free-energy contributions, thereby accounting for the direct influence of phonons on the elastic response. Surprisingly, the TDEC (QSA) showed a more realistic C$_{ij}$(T) trend compared to TDEC (QHA). Specifically, the QSA predicts a smooth and monotonic softening of all elastic constants with increasing temperature, while preserving the mechanical stability of the LCS over the entire temperature range considered. In contrast, the QHA predicts an anomalous increase in C$_{11}$(T), whereas C$_{12}$(T) and C$_{44}$(T) decrease continuously. With C$_{44}$(T) achieving a negative value at approximately T=600 K, it violates the Born mechanical stability criterion, suggesting mechanical instability. Also, as no structural instability or phase transition has been experimentally or theoretically reported for the LCS in this temperature range, these anomalous QHA results are doubtful for further analysis of temperature-dependent elastic moduli (TDEM) and other parameters. Most importantly, we therefore perform $\kappa_l$ calculation from 'Slack model+TDEC' with TDEC (QSA). The TDEM calculated within the QSA are presented in Figure \ref{moduli-QSA}. The B, Y and G decrease monotonically with increasing temperature, indicating the progressive weakening of inter-atomic bonding due to thermal expansion. In contrast, $\nu$ exhibits a slight increase, reflecting enhanced ductility as temperature elevates. Similarly, the sound velocities and $\rho$ show a marginal reduction with increasing temperature, primarily as a consequence of thermal expansion. The smooth evolution of these properties further confirms the mechanical stability of the material over the investigated temperature range.
\par Here, the vibrational thermodynamic quantities such as change in vibrational internal energy (E$_{vib}$), Gibbs free energy (G$_{vib}$), entropy (S$_{vib}$) and volume thermal expansion coefficient (C$_{vol}$) are determined primarily by the phonon frequencies through the vibrational Helmholtz free energy (F$_{vib}$) (see equation \ref{p}). 
\begin{equation}
	\begin{split}
		F^{vib}(X,T)=\frac{1}{2N}\sum_{\textbf{q},\nu}\hbar\omega(\textbf{q},\nu,X)\\
		+\frac{k_BT}{N}\sum_{\textbf{q},\nu}In\bigg[1-exp\bigg(\frac{-\hbar\omega(\textbf{q},\nu,X)}{k_BT}\bigg)\bigg]
	\end{split}
\label{p}	
\end{equation}
\par Since equilibrium volume undergoes minute changes with temperature, the phonon spectrum remains nearly unchanged in both QSA and QHA, leading to nearly similar values of these quantities (see figure \ref{dynamic1}(a-d)). In both cases, the C$_{vol}$ curves follow the Debye's T$^3$ law in lower temperature regime (T $<<$ $\Theta_D$): C$_{vol}$=$\frac{12}{5}$$\pi^4$nR$\big(\frac{T}{\Theta_D}\big)^3$; while at elevated temperature (T $>>$ $\Theta_D$), C$_{vol}$ tends to Dulong-Petit limit: C$_{vol}$=3nR, and obeys the classical thermodynamic law. In contrast, the thermal expansion coefficient ($\beta$) and constant-pressure heat capacity (C$_p$) rely on the temperature-dependent equilibrium volume and elastic response. Within the QHA, the F$_{vib}$ is included in the evaluation of elastic constants, which substantially softens the elastic response, particularly the C$_{44}$. Consequently, the softening modifies the bulk modulus (shown in figure \ref{dynamic2}), leading to enhanced $\beta$ through $\beta$=$\frac{\gamma C_{vol}}{BVol}$, where '$Vol$' is volume (refer to figure \ref{dynamic1}(f)). Also, this lead to larger C$_p$ according to C$_p$=C$_{vol}$+TVB$\beta^2$ as shown in figure \ref{dynamic1}(e).    
\subsection{Thermoelectric Properties}
\label{Thermoelectric}
The thermoelectric performance of the investigated LCS compound was evaluated by combining first-principles electronic structure calculations with semi-classical Boltzmann transport theory using Boltztrap2 code \cite{Madsen2006b} based on the GGA, GGA+SOC and HSE06. The carrier relaxation time ($\tau$) was estimated using the Deformation Potential (DP) theory by explicitly considering carrier effective masses, elastic moduli, and deformation potential constants \cite{Herring1956,Qin2017}. The lattice thermal conductivity was determined from the conventional Slack model, its temperature-dependent extension employing TDEC-derived elastic properties, and an ML model. These quantities were combined to evaluate the temperature-dependent thermoelectric figure of merit (ZT). To provide a comprehensive assessment, the transport properties were analyzed within both the constant relaxation-time approximation (CRTA) and the relaxation-time approximation (RTA), where the latter explicitly incorporates the carrier relaxation times obtained from the DP theory.
\begin{figure}[hbt!]
	\centering
	\includegraphics[width=8.5cm]{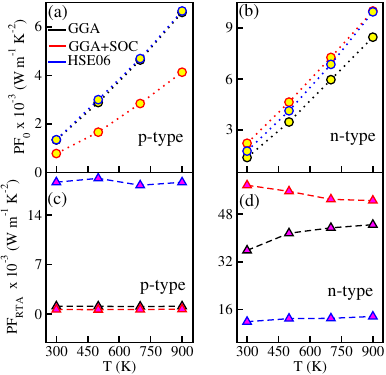}
	\caption{\label{Thermo-GGA} Variation of power factor (PF) with respect to temperature for p-type and n-type doping calculated in CRTA (a, b) and RTA (c, d), estimated from GGA, GGA+SOC, and HSE06 functionals.}
\end{figure}
\par From the transport distribution function given in equation \ref{T1}, the transport coefficients are calculated \cite{Madsen2006b}, 
\begin{equation}
	f(E, T)=\int \nu_k \otimes \nu_k \tau_k \delta(E-E_k) \frac{dk}{8\pi^3}
	\label{T1}
\end{equation}
Here, $\nu_k$ is the group velocity component, $E_k$ is the electronic state energy, and $\tau_k$ is the relaxation time.
\par Now, Fermi-Dirac distribution moments are, 
\begin{equation}
	\mathcal{L}^{(\alpha)}=e^2\int f(E, T) (E-\mu)^\alpha \bigg(-\frac{\delta f_0(E,\mu,T)}{\delta E}\bigg) dE
	\label{T2}
\end{equation}
\par From equation \ref{T2}, the electronic contribution to transport properties including electrical conductivity ($\sigma$), Seebeck coefficient (S), and electronic contribution to lattice thermal conductivity ($\kappa_e$), are evaluated as,
\begin{equation}
	\begin{split}
		\sigma = \mathcal{L}^{(0)} \\
		S = \frac{1}{eT}\frac{\mathcal{L}^{(1)}}{\mathcal{L}^{(0)}} \\
		\kappa_e = \frac{1}{e^2T}\bigg[\frac{\mathcal{L}^{(1)2}}{\mathcal{L}^{(0)}}-\mathcal{L}^{(2)}\bigg]
	\end{split}
	\label{a}
\end{equation}
\par To determine the lattice thermal conductivity ($\kappa_l$), the standard Slack model \cite{Slack1987} given in equation \ref{T3}, the modification with TDEC and ML from equation \ref{eq1} were employed,
\begin{equation}
	\kappa_l=\frac{A\bar{M}\Theta_D^3\delta}{\gamma^2Tn^{2/3}}
	\label{T3}
\end{equation}
Here, A=$\frac{2.43 x 10^{-8}}{1-0.514/\gamma+0.228/\gamma^2}$, $\bar{M}$ average atomic mass, $\Theta_D$ the Debye temperature, $\delta^3$ volume per atom, $\gamma$ the Gruneisen parameter, and n number of atoms per unit cell. 
\begin{figure}[t!]
	\centering
	\includegraphics[width=6.5cm]{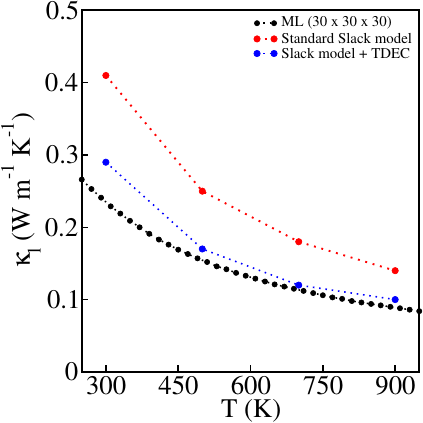}
	\caption{\label{Kl} The lattice thermal conductivity (K$_l$) calculated from Slack model, Slack model+TDEC and ML.}
\end{figure}
\begin{figure*}[htb!]
	\centering
	\includegraphics[width=17cm]{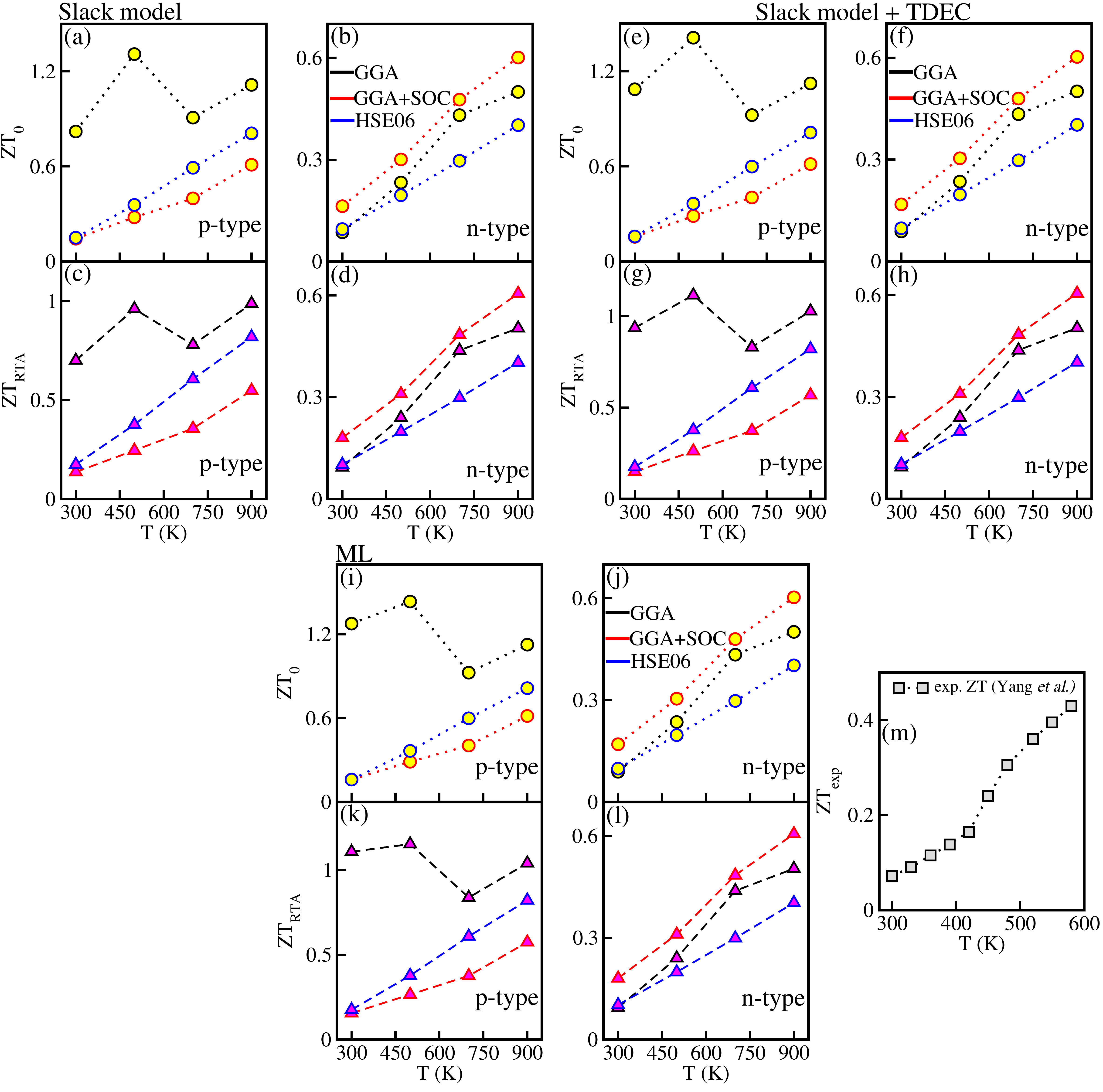}
	\caption{\label{ZT} Thermoelectric figure of merit (ZT) from CRTA and RTA using K$_l$ derived from (a-d) standard Slack model, (e-h) Slack model+TDEC, and (i-l) ML based on GGA, GGA+SOC and HSE06. (m) the available experimental ZT for LCS compound from Yang \textit{et al} \cite{Yang2022}.}
\end{figure*}
\par The electronic transport properties corresponding to chemical potential ($\mu$) presented in figures S1-S3, revealed the n-type semiconducting nature of the investigated LSC compound, with electrons as majority charge carriers. The electronic band-splitting due to the implementation of SOC modifies the dispersion in the vicinity of the E$_F$, this led to changes in the band curvature and valley degeneracy. From the DOS plot shown in figure \ref{atomic-dos}, it is evident that the SOC DOS when compared to the standard GGA DOS undergoes enhanced intensity along the CB edge while reduces along the VB edge. This amplifies the PF$_0$ (PF obtained from CRTA) for n-type doping i.e., PF$_0$(GGA+SOC) $>$ PF$_0$(GGA). Conversely, the reduced DOS at the VB edge suppresses the transport performance under p-type doping, yielding PF$_0$(GGA) $>$ PF$_0$(GGA+SOC). Due to this reason, the electronic thermal conductivity ($\kappa_e$) which satisfies the Wiedemann-Franz relation \cite{Graf1996} i.e., $\kappa_e$=L$\sigma$T, where L is the Lorenz number, follows the electrical conductivity ($\sigma$) trend. Consequently, the induced SOC enhancement (suppression) of electron (hole) transport also leads to a corresponding increase (decrease) in $\kappa_e$ for n-type (p-type) carriers. Relative to the GGA and GGA+SOC calculations, the transport coefficients predicted from HSE06 functional are expected to yield a more reliable results due to accurate description of electronic structure by correcting the exchange-induced self-interaction error. The corresponding increase in E$_g$ enlarges the energy separation (E-$\mu$) term, entering the transport integrals equations \ref{T2} and \ref{a}, which enhances the Seebeck coefficient (S). Simultaneously, the reduced carrier concentration and transport distribution near the E$_F$ lead to a lower $\sigma$ value. Such an inverse correlation between S and $\sigma$ determines the resulting PF$_0$.
\par The carrier relaxation time ($\tau$) was estimated within the framework of the Bardeen-Shockley DP theory, where acoustic phonon scattering was considered as the dominant carrier scattering mechanism. The DP constant (E$_d$) was estimated from the linear shift of band-edge energy under small uniaxial strain (shown in figure S4). These quantities were used to evaluate the temperature-dependent relaxation time, which was subsequently employed in the Boltzmann transport calculations. For three dimensional semiconductor, relaxation time for electrons and holes ($\tau_e$ and $\tau_h$) are given by \cite{Qin2017}
\begin{equation}
	\tau_{e/h}=\frac{2\sqrt{2\pi}C\hbar^4}{3(k_BT)^(\frac{3}{2})(m^*_{e/d})^\frac{3}{2}E_d^2(CBM/VBM)}
	\label{z}
\end{equation} 
\par From the temperature-dependent normalized carrier relaxation time ($\tau$/$\tau_0$, where $\tau_0$=10$^{-14}$ s), shown in Figure S5, the T$^{-1}$ dependence of $\tau$ explains the monotonic decrease observed, consistent with the enhanced acoustic phonon scattering with elevated temperature. The distinct functional dependence of $\tau$ is evident from the comparison between the GGA and HSE06 estimated data, where the GGA predicts substantially longer $\tau_e$, and the HSE06 functional yields significantly larger $\tau_h$. This contrasting behaviour originates from the different electronic structures predicted by the two functionals, which modify the m$^*$ and E$_d$ entering equation \ref{z}. The larger $\tau_e$ within GGA is consistent with the enhanced n-type PF$_{RTA}$, while the pronounced increase in $\tau_h$ predicted by HSE06 partially compensates for its reduced electrical conductivity. Consequently, the thermoelectric response of the LCS compound is governed not only by the band structure but also by the strong functional dependence of the carrier scattering rates.
\par Figure \ref{ZT} compare the efficiency (ZT) obtained using the $\kappa_l$ predicted from the conventional Slack model, the temperature-dependent Slack model employing the elastic properties derived from TDEC and the ML model. The corresponding $\kappa_l$ are shown in Figure \ref{Kl}. Irrespective of the adopted model, $\kappa_l$ decreases monotonically with temperature, following the characteristic T$^{-1}$-like behaviour associated with the increasing anharmonic phonon-phonon Umklapp scattering. Moreover, the three approaches derive remarkably low $\kappa_l$ ($<$ 0.5 W m$^{-1}$ K$^{-1}$) throughout the considered temperature range. Interestingly, the obtained $\kappa_l$ from the Slack model+TDEC is closer to the ML model compared to the standard Slack model. From the ZT equation given below:
\begin{equation}
	ZT=\frac{S^2\sigma T}{\kappa_e+\kappa_l}
\end{equation}
Such an ultralow $\kappa_l$ considerably suppresses the denominator, rendering the thermoelectric performance predominantly controlled by the electronic transport coefficient. 
\par Within the CRTA, the GGA yields the largest p-type ZT, whereas the GGA+SOC produces the highest n-type ZT. This behaviour confirms the trend where SOC enhances the CB transport distribution while reducing the VB contribution. In contrast, the HSE06 predicts intermediate ZT values due to a more balanced derivation of electronic transport. The consideration of $\tau$ incorporated within the RTA reduced the magnitude of ZT relative to CRTA, where $\tau$ directly alters the balance between $\sigma$ and $\kappa_e$. To assess the reliability of the DFT approaches and models employed for $\kappa_l$ estimation, the theoretically obtained ZT are compared to an available experimental ZT (ZT$_{exp}$) from Yang \textit{et al} \cite{Yang2022}. Irrespective of the approaches and models employed, the ZT from n-type doping is underestimated compared to ZT$_{exp}$. For p-type doping, the $\kappa_l$ from the conventional Slack model systematically overestimates the ZT due to the simplified treatment of phonon scattering. The Slack+TDEC model provides a more realistic ZT description, improving the agreement with ZT$_{exp}$ although a moderate overestimation remains, indicating that higher-order anharmonic effects and complex phonon-phonon interactions are not fully captured. The ML-based $\kappa_l$ model further improves the prediction accuracy, reproducing near ZT$_{exp}$. This highlights the critical role of accurate lattice thermal transport modelling in quantitative thermoelectric predictions. Moreover, the electronic structure treatment also affects the calculated ZT. Overall, the agreement with experiment follows the trend: 
\par ML + HSE06 $>$ Slack+TDEC + HSE06 $>$ Slack + HSE06. 
\section{Conclusion}
In summary, we investigate the thermoelectric properties of LiCdSb using first-principles calculations combined with semi-classical Boltzmann transport theory, incorporating the ML and Slack+TDEC for $\kappa_l$ estimation. To capture more reliable electronic transport coefficients, the HSE06 hybrid functional was adopted in addition to the GGA and GGA+SOC. While the conventional Slack model overestimates the thermoelectric performance, both Slack+TDEC and ML-based $\kappa_l$ yield ZT values in close agreement with experiment, with the ML approach providing the best overall agreement. These results demonstrate that accurate prediction of thermoelectric performance requires simultaneous improvement of both electronic structure and lattice thermal transport. The combined HSE06 and advanced $\kappa_l$ framework, particularly the ML-assisted approach, provides a reliable and computationally efficient strategy for quantitative prediction and design of thermoelectric materials.

\section{Acknowledgments}
R. Zosiamliana gratefully acknowledges the assistance received from the "National Fellowship and Scholarship for Higher Education of Scheduled Tribe Students (NFST), Ministry of Tribal Affairs, Government of India". Award No.: 202223-NFST-MIZ-01557 (dated 28.06.2023).\\
\par Lalhriat Zuala acknowledges support for this research work provided by Anusandhan National Research Foundation (ANRF), Government of India, vide Grant No.: ANRF/IRG/ 2024/000695/PS (dated 22.3.2025).\\

\textbf{A. Laref} acknowledges support from the "Research Center of the Female Scientific and Medical Colleges",  Deanship of Scientific Research, King Saud University.

\nocite{*}
\bibliographystyle{elsarticle-num}
\bibliography{LiCdSb}

\end{document}